\documentclass[12pt]{iopart}

\usepackage{iopams}  
\usepackage{color}
\usepackage{caption}
\usepackage{graphicx}
\begin{document}

\title{Concept of SUb-atmospheric Radio-frequency Engine (SURE) for near space environment}
\author{Xiaogang Yuan$^{1, 2}$, Lei Chang$^{1, 2*}$, Xinyue Hu$^3$, Xin Yang$^1$, Haishan Zhou$^{1, 2}$, and Guangnan Luo$^{1, 2}$}
\address{$^1$Institute of Plasma Physics, Chinese Academy of Sciences, Hefei 230031, China}
\address{$^2$Graduate Island of Sciences, University of Science and Technology of China, Hefei 230026, China}
\address{$^3$School of Aeronautics and Astronautics, Sichuan University, Chengdu, 610065, China}
\ead{lei.chang@ipp.ac.cn}

\begin{abstract}
A concept of SUb-atmospheric Radio-frequency Engine (SURE) designed for near space environment is reported. The antenna wrapping quartz tube consists of two solenoid coils with variable separation distance, and is driven by radio-frequency power supply ($13.56$~MHz-$1$~kW). The discharge involves inductive coupling under each solenoid coil and capacitive coupling between them. This novel scheme can ionize the filling air efficiently for the entire pressure range of $32\sim 5332$~Pa in near space. The formed plasma density and temperature are up to $2.23\times 10^{18}~\textrm{m}^{-3}$ and $2.79$~eV, respectively. The influences of separation distance, input power, filling pressure and the number of solenoid turns on discharge are presented in detail. This air-breathing electric propulsion system has no plasma-facing electrode and does not require external magnetic field, and is thereby durable and structurally compact and light. 
\end{abstract}

\maketitle

\twocolumn
Near space, with typical altitude from $20$~km (Armstrong line) to $100$~km (Karman line) and pressure in range of $32\sim 5332$~Pa, lies in the region between conventional aviation and aerospace\cite{Chamberlain:1987aa, Woolley:2009aa, Xu:2011aa}. It has been attracting great attention recently for civil and military explorations, owing to certain advantages such as persistent intelligence, surveillance, reconnaissance, long endurance, beyond line of sight communication, and low cost access to space-like performance\cite{Knoedler:2005aa, Airforce:2005aa, Moomey:2005aa, Schmidt:2008aa, Young:2009aa, Wang:2011aa, Wang:2014aa}. Propulsion system plays the most critical role for high-altitude aircraft and low-altitude satellite in this space, in terms of long-duration flight and gesture control\cite{Schmidt:2008aa, Young:2009aa}. However, conventional aeronautic propulsion schemes including turbine and propeller cannot work efficiently because of low density atmosphere, and present astronautic propulsion schemes such as electric thrusters are neither suitable due to high power consumption for ionization and excessive aerodynamic drag\cite{Marcus:2009aa, Jonathan:2009aa, Charles:2009aa, Lev:2019aa}. Although there exist various atmospheric discharge methods\cite{Winter:2015aa}, such as corona discharge\cite{Fridman:2005aa}, dielectric barrier discharge\cite{Kogelschatz:2003aa, Corke:2010aa}, atmospheric pressure plasma jet (APPJ)\cite{Schutze:1998aa, Jeong:1998aa, Babayan:1998aa, Herrmann:1999aa, Park:2001aa}, cold plasma torch\cite{Koinuma:1992aa}, glow discharge\cite{Roth:1995aa, Massines:2003aa}, microhollow cathode discharge\cite{Stark:1999aa}, and surface-wave discharge\cite{Moisan:1998aa}, most of them involve plasma-facing electrode which limits the lifespan due to erosion problem and the efficiency due to formed sheath\cite{Lieberman:2005aa}. This letter reports a novel concept of RF plasma thruster (named SURE for SUb-atmospheric Radio-frequency Engine) which can ionize the filling air efficiently for the entire pressure range of $32\sim 5332$~Pa in near space. There is no plasma-facing electrode, yielding long life and high efficiency, and does not require external magnetic field so that can be structurally compact and light. Moreover, it is an air-breathing system that does not need to bring propellant as traditional electric thrusters do in aerospace, thus can further extend the lifespan. Hence, this concept is a very promising candidate for near-space propulsion. 

The SURE experiment includes three parts: plasma source, pumping system and diagnostics. The source consists of a cylindrical quartz tube in length of $0.6$~m and diameter of $0.05$~m, RF power supply of frequency $13.56$~MHz and maximum power $1$~kW, impedance matching network, and an innovative antenna wrapping the quartz tube. This antenna comprises two legs which are solenoid coils, separated by a variable distance to introduce voltage difference for acceleration, and connected to the high-voltage output and ground of power supply respectively. The inductive coupling of each leg can be controlled separately by adjusting the number of solenoid turns, and the capacitive coupling and acceleration between two legs are controllable through changing the separation distance and applied voltage. Please note that this structure is different from the antenna of two ring electrodes which only involves capacitive coupling effect\cite{Winter:2015aa}. The pumping system includes gas cylinder, mass flow controller, vacuum meter and mechanical pump. The filling pressure inside the quartz tube can be varied from $0.1$~Pa to $1$~atm. The diagnostics include RF-compensated Langmuir probe, which provides the information of plasma density and temperature via the measured I-V curve, and digital camera for imaging and video recording. The probe tip is made of Tungsten and $1.5\times 10^{-5}$~m in diameter and $0.01$~m long, and sits on axis and $\sim 0.03$~m away from the ground leg of antenna. The diagnostic data are collected and analyzed by an acquisition system.

\begin{figure}[ht]
\begin{center}$
\begin{array}{l}
(a_1)\\
\hspace{0.6cm}\includegraphics[width=0.42\textwidth,angle=0]{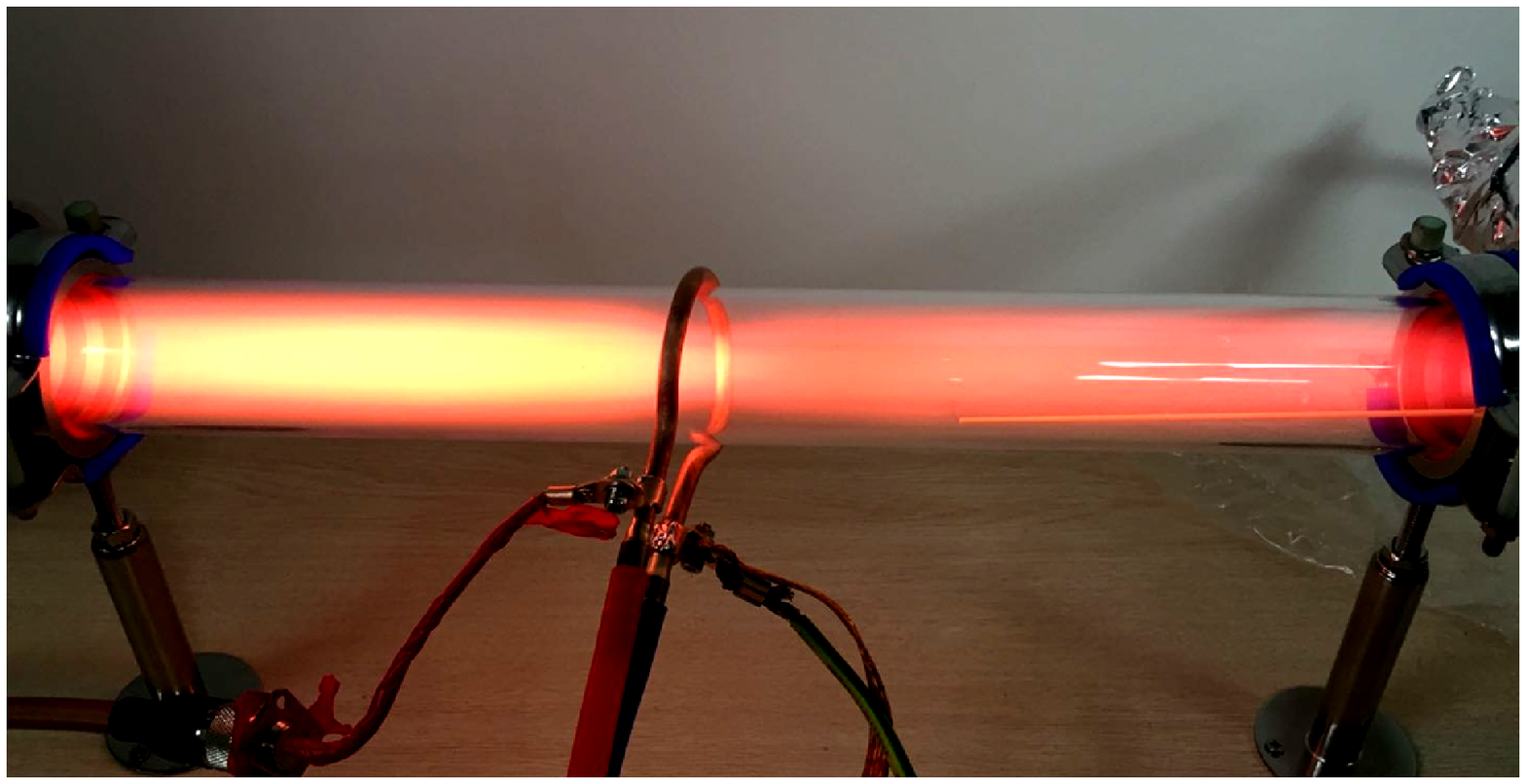}\\
(a_2)\\
\hspace{0.6cm}\includegraphics[width=0.42\textwidth,angle=0]{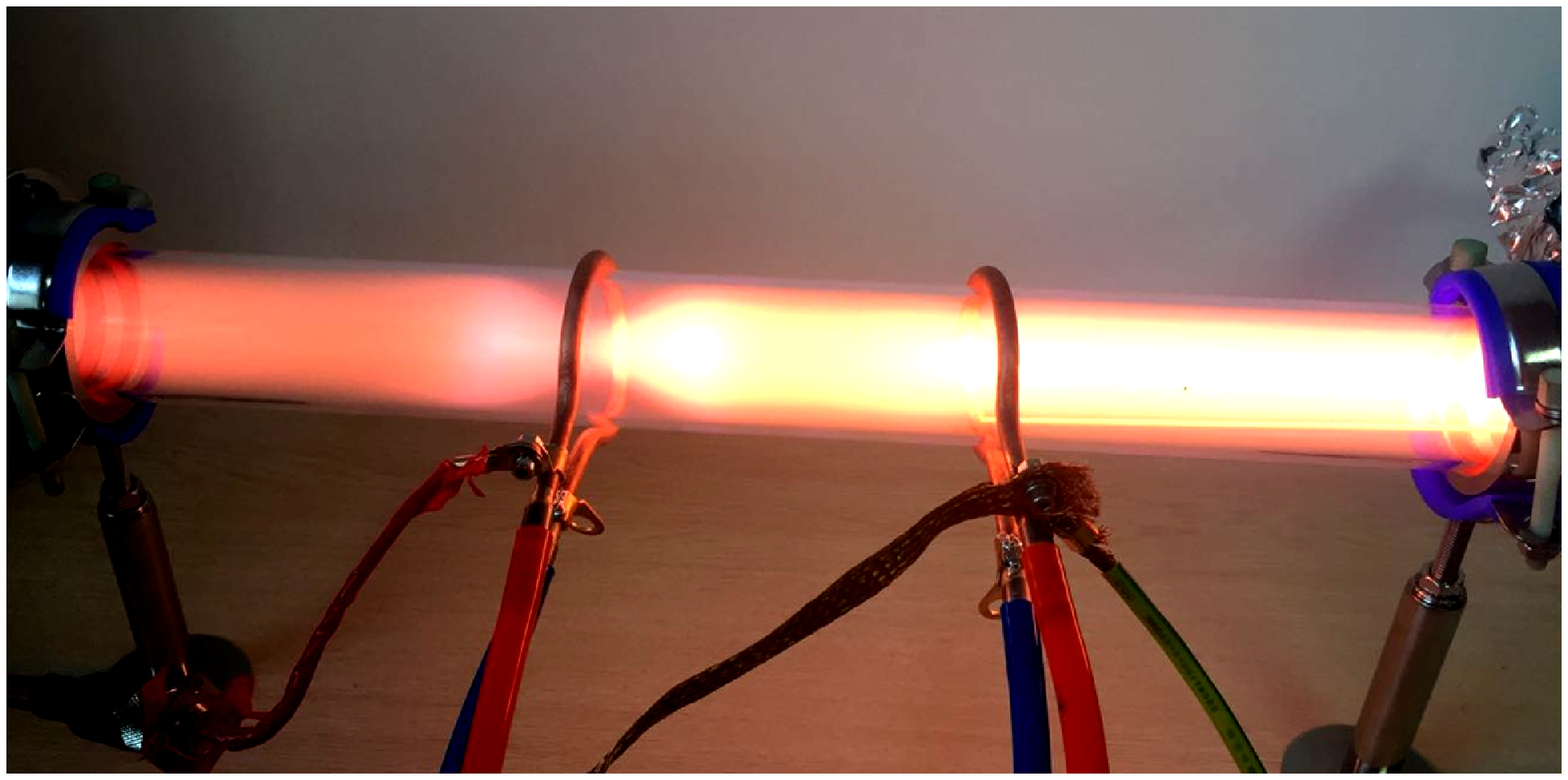}\\
(b_1)\\
\hspace{0.6cm}\includegraphics[width=0.42\textwidth,angle=0]{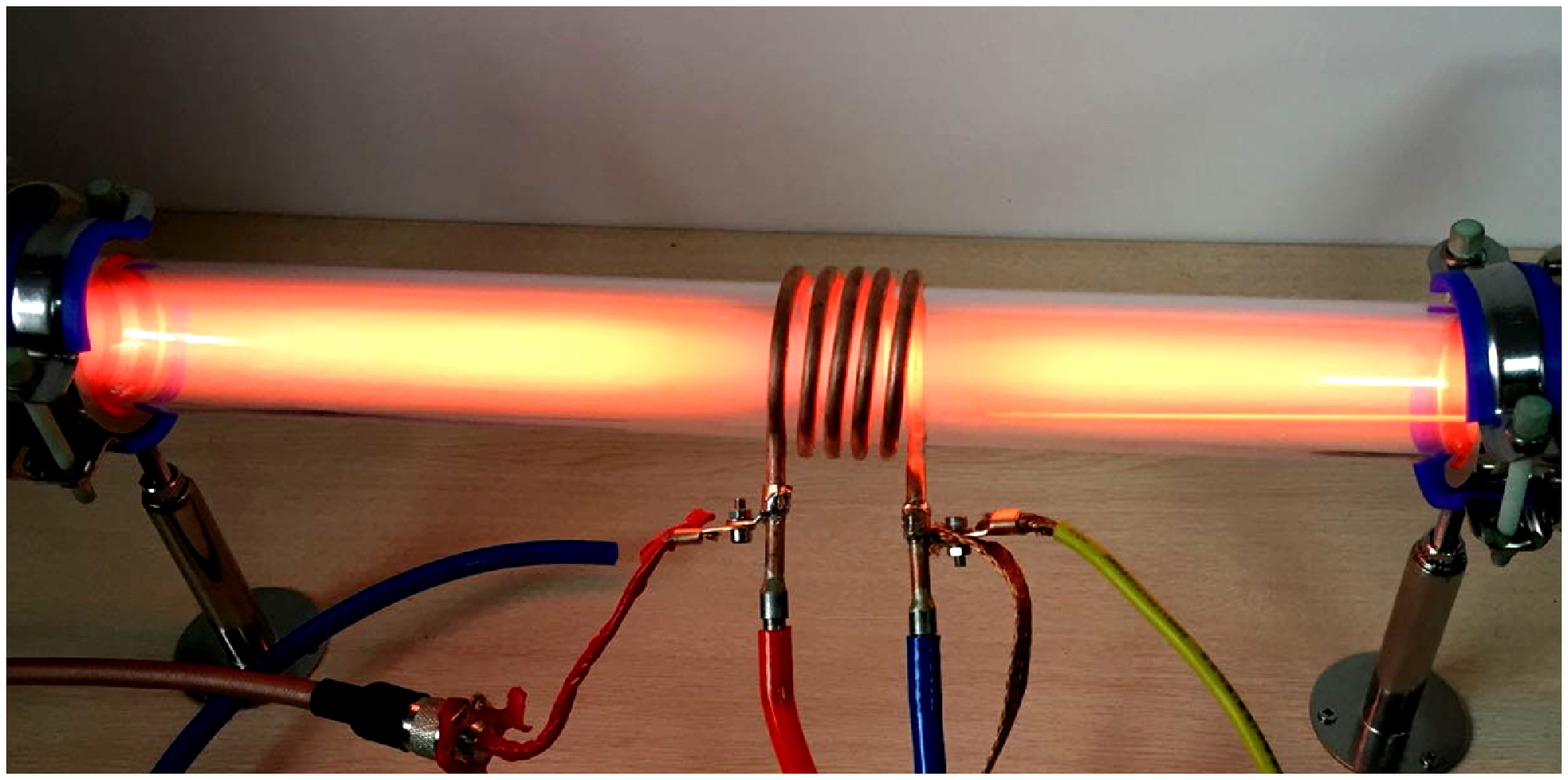}\\
(b_2)\\
\hspace{0.6cm}\includegraphics[width=0.42\textwidth,angle=0]{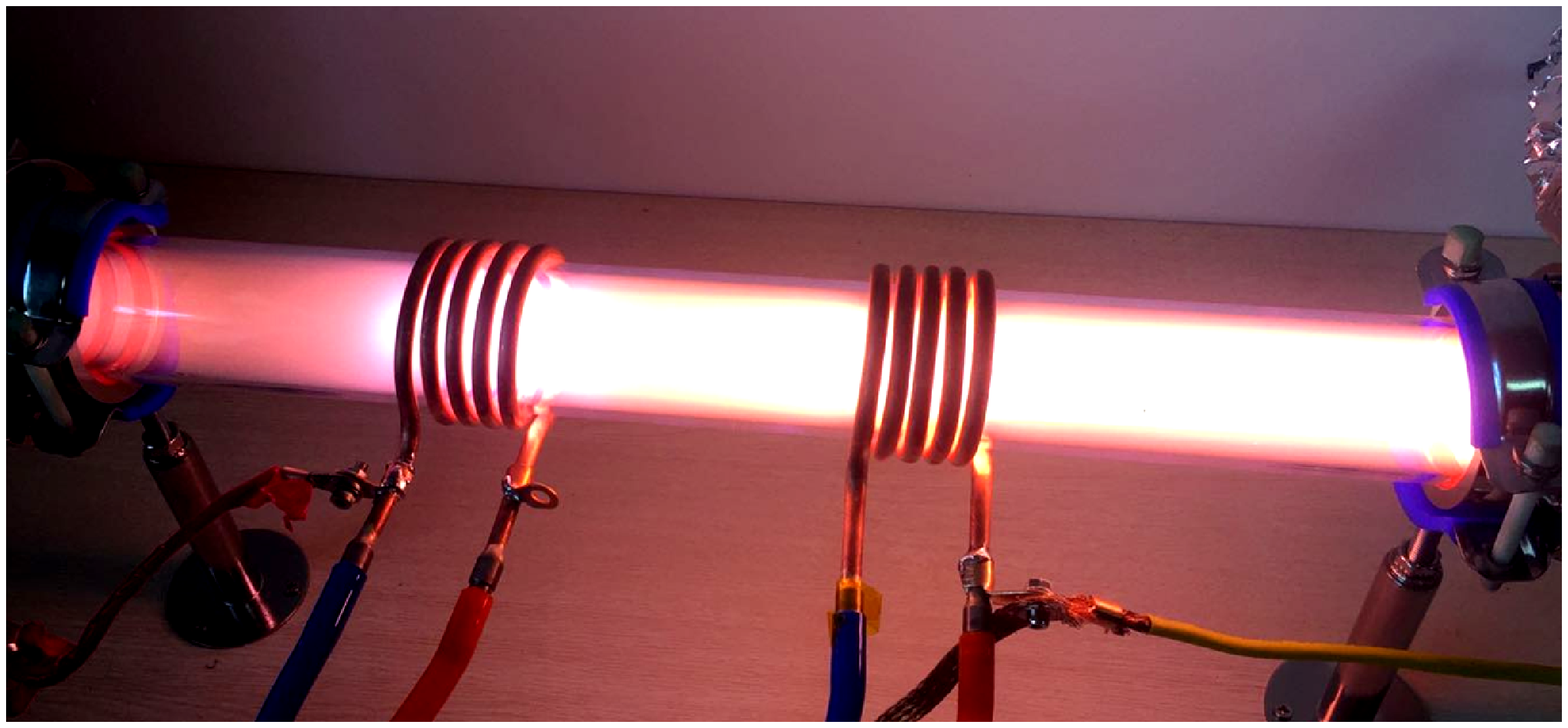}\\
\end{array}$
\end{center}
\caption{Comparison between single and dual antennas for filling pressure of $200$~Pa and input power around $800$~W.}
\label{fg1}
\end{figure}

\begin{figure}[ht]
\begin{center}$
\begin{array}{l}
(a)\\
\hspace{0.8cm}\includegraphics[width=0.4\textwidth,angle=0]{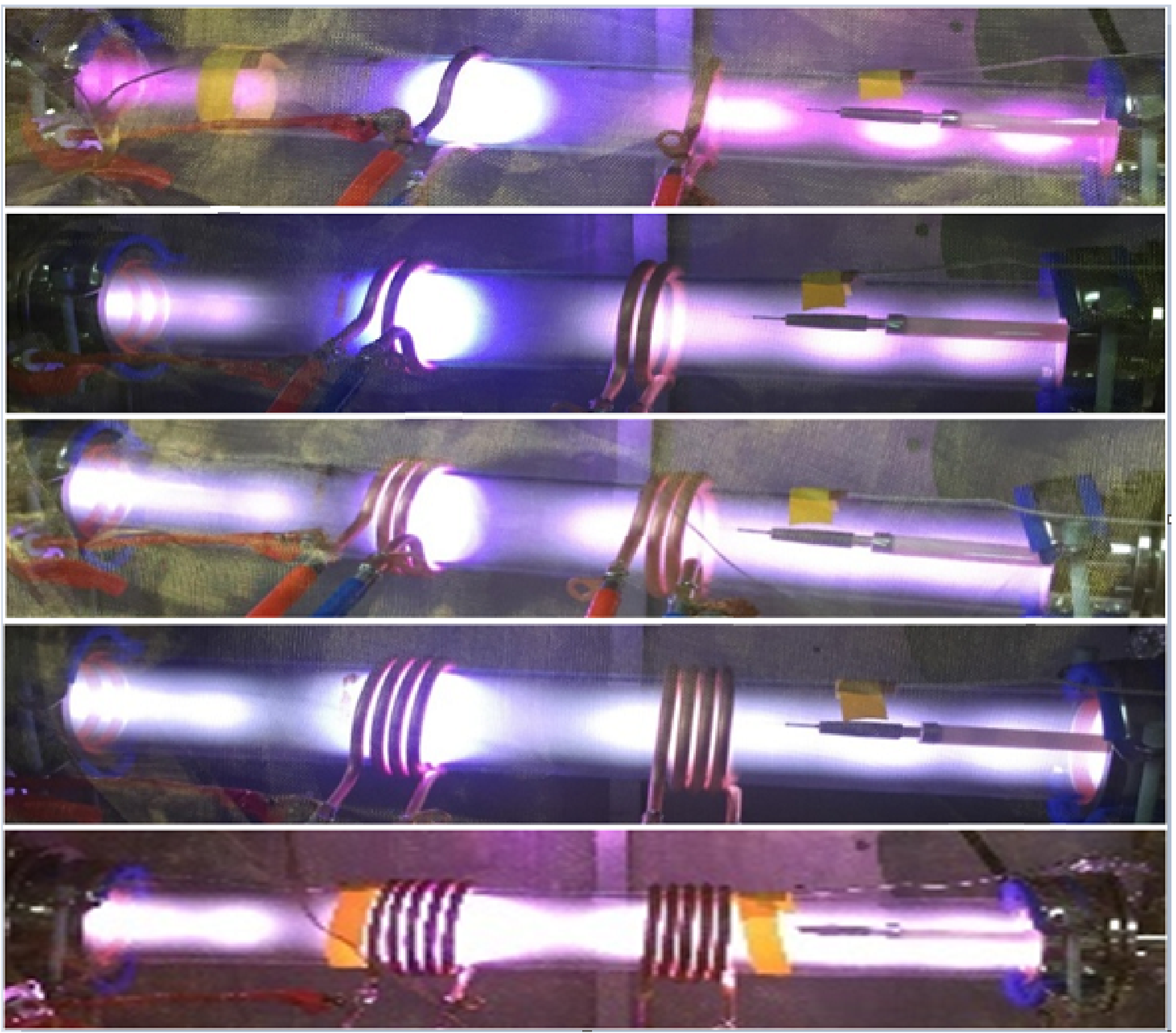}\\
\vspace{-0.5cm}(b)\\
\hspace{-0.15cm}\vspace{-0.3cm}\includegraphics[width=0.5\textwidth,angle=0]{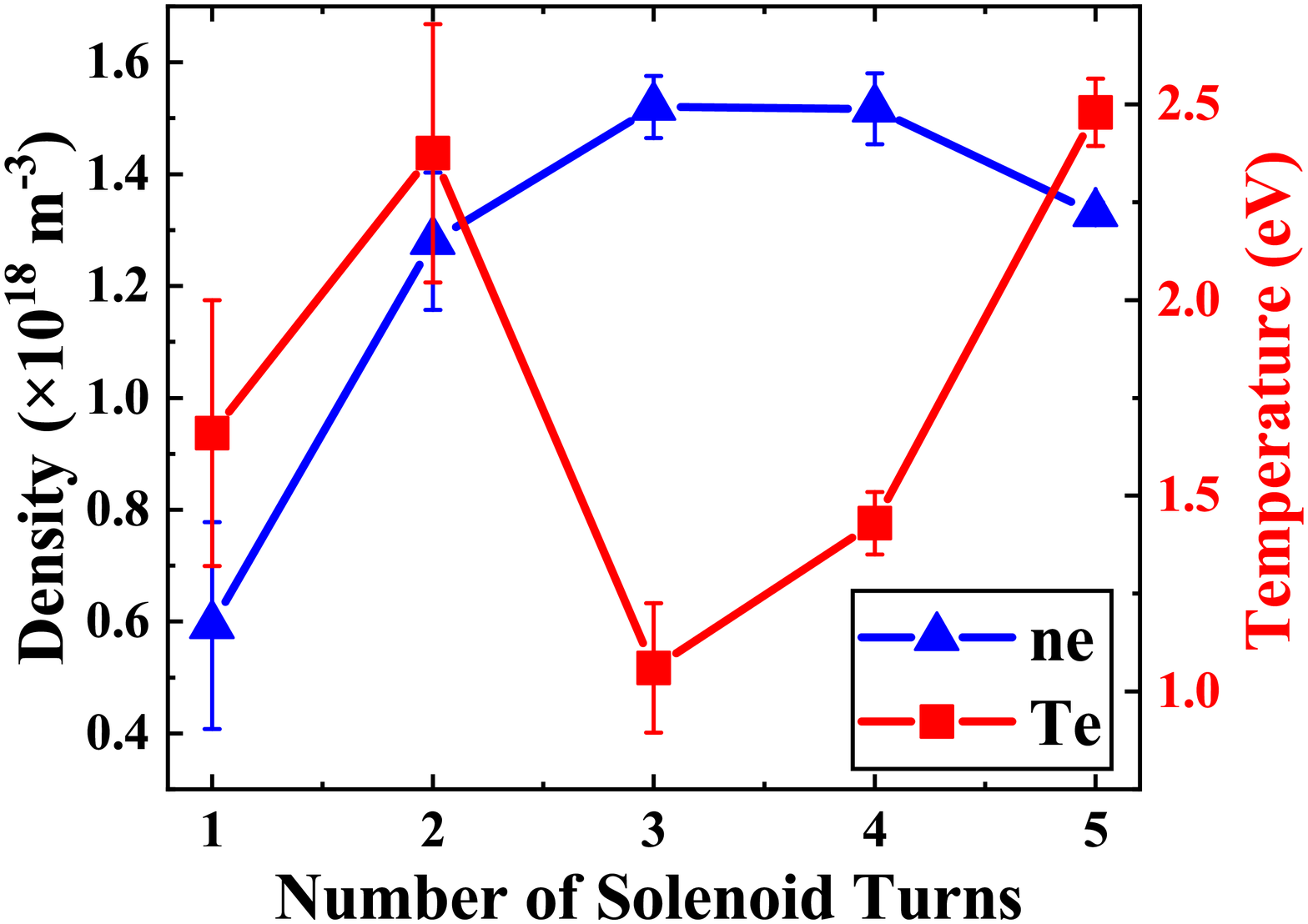}
\end{array}$
\end{center}
\caption{The influence of the number of solenoid turns on discharge for input power of $800$~W and filling pressure of $200$~Pa.}
\label{fg2}
\end{figure}

\begin{figure}[ht]
\begin{center}$
\begin{array}{l}
\vspace{-0.5cm}(a)\\
\hspace{-0.4cm}\includegraphics[width=0.55\textwidth,angle=0]{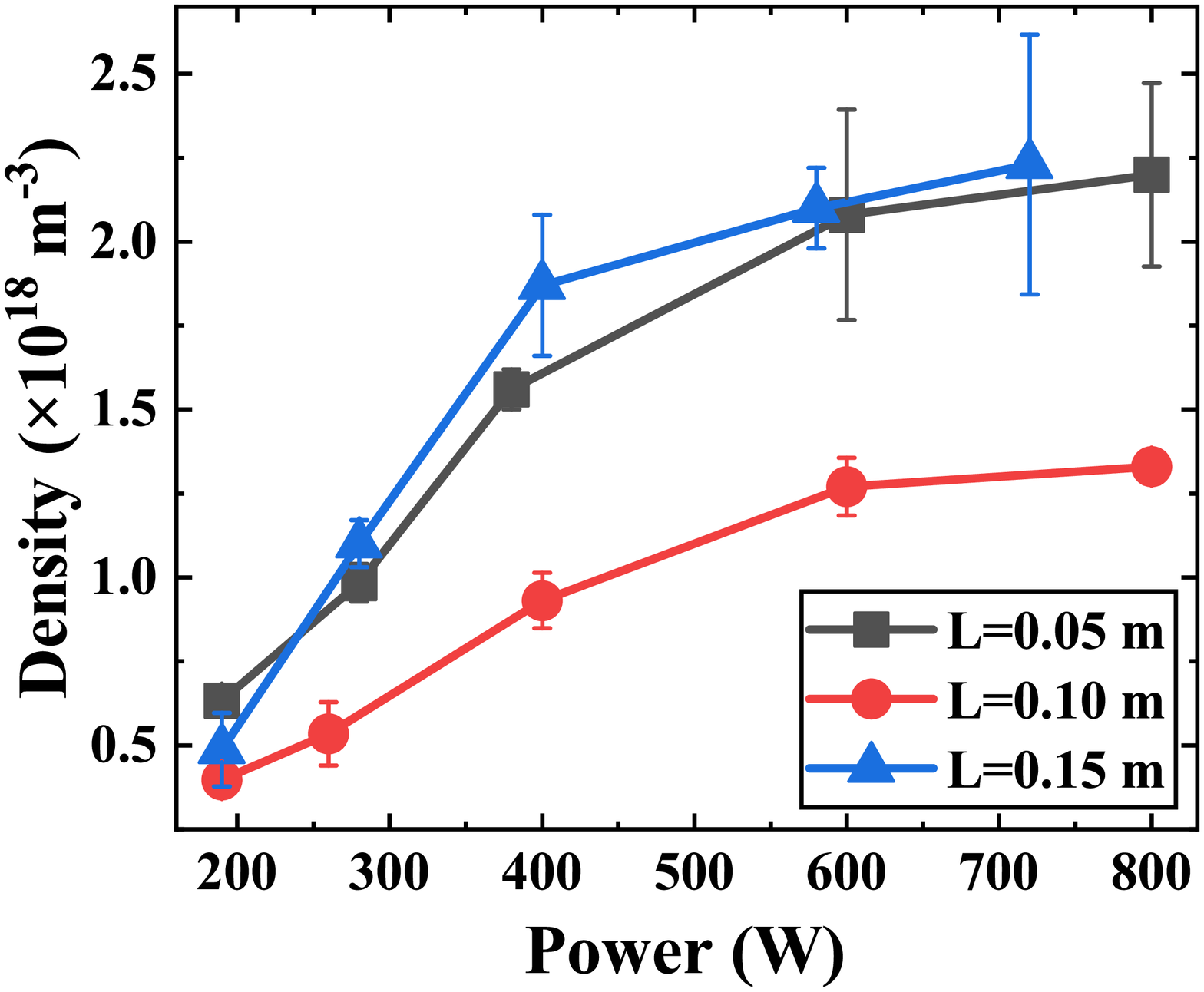}\\
\vspace{-0.5cm}(b)\\
\hspace{-0.4cm}\includegraphics[width=0.55\textwidth,angle=0]{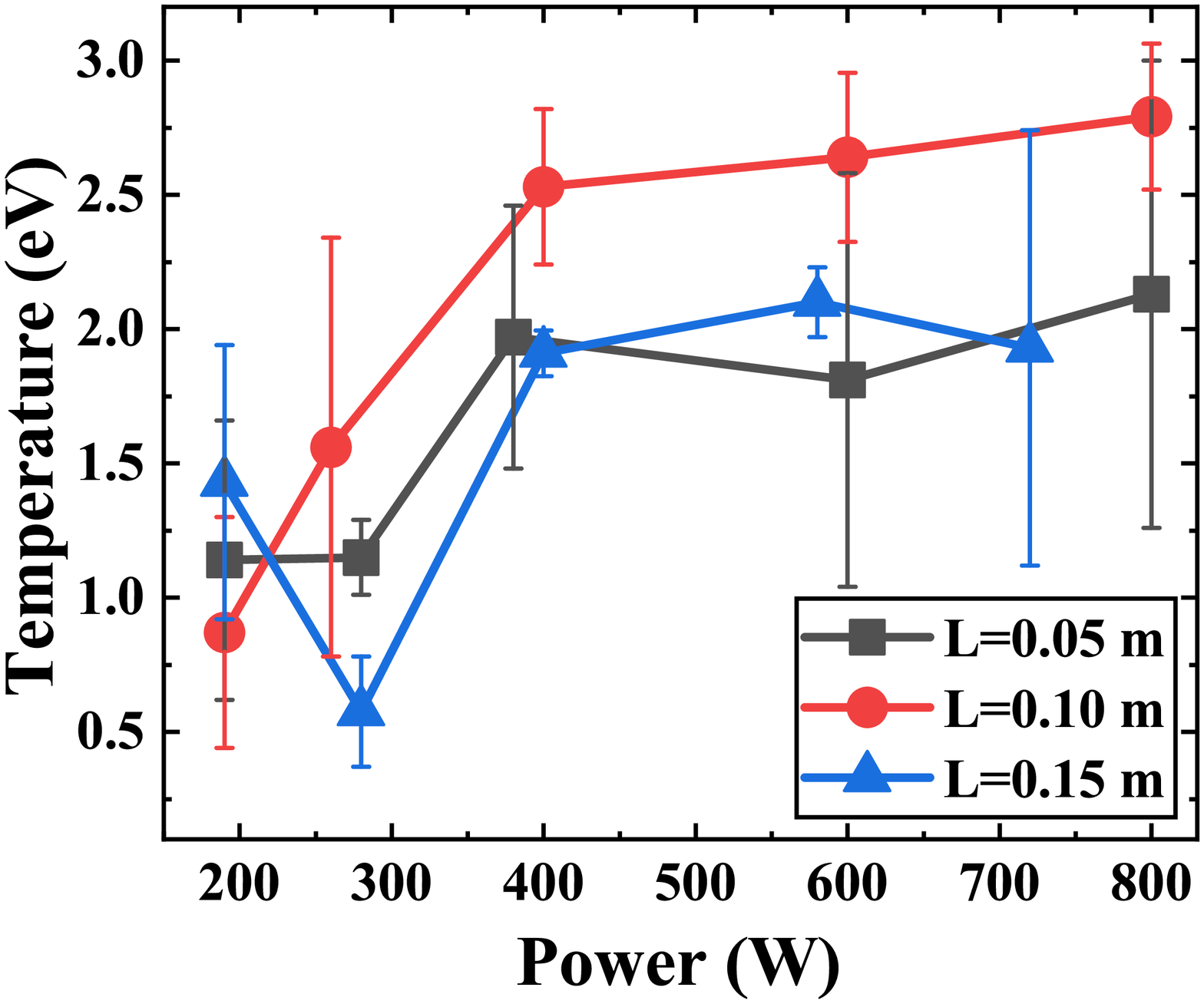}
\end{array}$
\end{center}
\caption{Dependence of plasma density (a) and temperature (b) on input power and separation distance between antenna legs ($5$ turns) for filling pressure of $200$~Pa.}
\label{fg3}
\end{figure}

\begin{figure}[ht]
\begin{center}
\hspace{-0.6cm}\includegraphics[width=0.52\textwidth,angle=0]{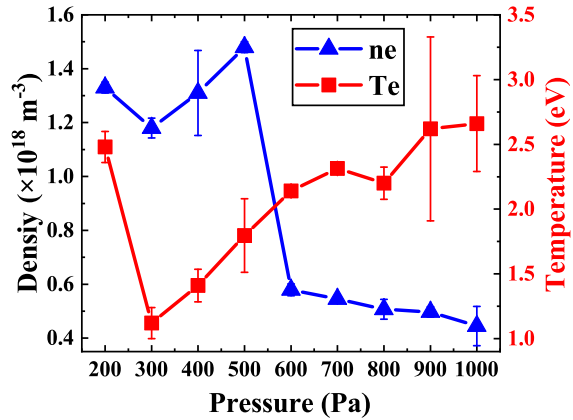}
\end{center}
\caption{Dependence of plasma density and temperature on filling pressure for dual antennas with $5$ turns, separation distance of $0.1$~m and input power of $800$~W.}
\label{fg4}
\end{figure}

\begin{figure}[ht]
\begin{center}
\includegraphics[width=0.47\textwidth,angle=0]{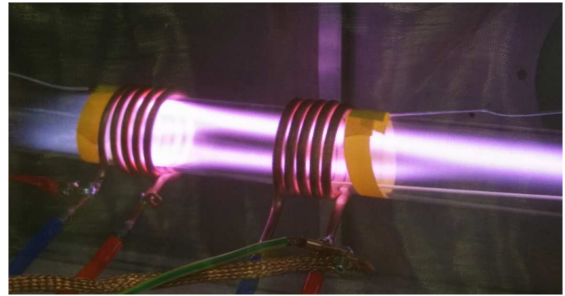}
\end{center}
\caption{Typical image of SURE discharge at high pressure ($3500$~Pa) for $800$~W.}
\label{fg5}
\end{figure}

Figure~\ref{fg1} shows the typical discharges driven by single (one-leg) and dual (two-leg) antennas. 
Examples include $1$ turn and $5$ turns of solenoid coils. The filling pressure is $200$~Pa and the input power is around $800$~W. We can see that the discharge for dual antennas is much brighter than that for single antenna, implying higher plasma density; actually, the measured plasma density is nearly two-order higher: ($b_1$) $3.99\times 10^{16}~\textrm{m}^{-3}$ for $890$~W ($2.94$~eV) and ($b_2$) $1.33\times 10^{18}~\textrm{m}^{-3}$ for $800$~W ($2.48$~eV). Moreover, the discharge has axial preference regarding the middle of dual antennas, which labels the acceleration feature of the second leg. Please note that here the working medium is air, to demonstrate the superiority of SURE; in following sections, it is changed to argon for easy diagnostics and clear parameter study. Detailed influence of the number of solenoid turns on discharge for dual antennas is given in Fig.~\ref{fg2} for input power of $800$~W and filling pressure of $200$~Pa.
It can be seen that for increased number of solenoid turns, the discharge is brighter and more continuous, i. e. from bullets to column; moreover, the plasma density increases first and then decreases slightly, maximizing around $3$ turns where the plasma temperature minimizes. These different variations of plasma density and temperature, i. e. when the number of solenoid turns is bigger than $3$ (including), may imply that the ionization procedure which generates plasma is opposite to the heating procedure that accelerates charged particles to high temperature. Figure~\ref{fg3} illustrates the dependence of plasma density and temperature on input power and the separation distance between antenna legs ($5$ turns) for $200$~Pa. 
We can see that both plasma density and temperature increase for increased power, which is reasonable as expected. Interestingly, the plasma density is highest for separation distance of $0.15$~m and lowest for separation distance of $0.1$~m, whereas the plasma temperature is on the contrary. This implies again that the ionization procedure is opposite to the heating procedure, and there may be a competition between them. The maximum plasma density and temperature are $2.23\times 10^{18}~\textrm{m}^{-3}$ and $2.79$~eV, respectively. This competition hypothesis gains further evidence from the dependence of plasma density and temperature on filling pressure above $500$~Pa, shown in Fig.~\ref{fg4}. Results are for dual antennas with $5$ turns, separation distance of $0.1$~m and input power of $800$~W. 
One can see that the plasma density largely decreases for enhanced pressure, while the plasma temperature increases on the whole. These opposite trends of plasma density and temperature are consistent with previous studies\cite{Anjum:2020aa, Hu:2020aa}. The underlying physics may be correlated with the mean free path of electrons, which is short for high density and thereby shortens the acceleration path, yielding low temperature under certain conditions. The monotonicity breaks at $300$~Pa and the plasma density drops sharply for pressure increased from $500$~Pa to $600$~Pa, implying that the discharge may turn into a different mode there. Indeed, when we further increase the pressure from $1000$~Pa to $5500$~Pa (for both argon and air), the discharge becomes bifurcate, filamentous and quivering (videos available online), which means that the mode transits from $\alpha$ to $\gamma$ according to the typical features of atmospheric RF discharge\cite{Park:2001aa, Godyak:1986aa, Vidaud:1988aa, Raizer:1995aa, Liu:2009aa}. Figure~\ref{fg5} gives a typical image of discharge at high pressure ($3500$~Pa) from which we can see bifurcated plasma striations. 

In summary, to provide an efficient propulsion scheme for near-space applications, the authors invent a novel radio-frequency plasma thruster (SURE) which can work steadily for the entire pressure range of $32\sim 5332$~Pa in near space. The special antenna has two legs, separated by a controllable distance, and each leg is a solenoid coil. This structure can incorporate capacitive coupling and inductive coupling simultaneously, and enhance the ionization and acceleration capabilities. For RF power supply of frequency $13.56$~MHz and power $1$~kW, the formed plasma density and temperature can be as high as $2.23\times 10^{18}~\textrm{m}^{-3}$ and $2.79$~eV, respectively. It is found that the variation of plasma density is opposite to that of plasma temperature with the separation distance, filling pressure and the number of solenoid turns, although they both increase with enhanced power as expected. This implies that the ionization and heating procedures are opposite, and may be correlated with the mean free path of electrons, which is short for high density and thereby shortens the acceleration path, leading to low temperature. Further experiment will be devoted progressively to the design of SURE prototype, performance test in a vacuum chamber ($190\sim 270$~K \& $32\sim 5332$~Pa) by measuring the thrust and specific impulse, and flight demonstration in near space based on high-altitude balloon or dirigible. 

This work is supported by Chinese Academy of Sciences (CAS) ``100 Talent" and Users with Excellence Project of Hefei Science Center CAS (2018HSC-UE006). 

\section*{Data Availability}
The data that support the findings of this study are available from the corresponding author upon reasonable request.\\

\section*{References}
\bibliographystyle{unsrt}

\end{document}